\documentclass{amsart}

\usepackage{amssymb}
\usepackage[mathscr]{eucal}
\usepackage{epic}
\usepackage{eepic}

\newtheorem{conjecture}{Conjecture}
\newcommand{\ds}{\displaystyle}
\def\EXP{\textrm{{\Large e}}}
\newcommand{\ii}{\mathsf{i}}

\begin{document}

\vspace{2cm}

\title[]{Lattice system of interacting spins in the thermodynamical limit}%
\author{S. Sergeev}%
\address{Department of Theoretical Physics,
Research School of Physical Sciences and Engineering, Australian
National University, Canberra ACT 0200, Australia}
\email{sergey.sergeev@anu.edu.au}

\thanks{This work was supported by the Australian Research Council}%


%-----------------------------------------------------
\begin{abstract}
In this paper we investigate some particular spin lattice (a
higher dimensional generalization of a spin chain) related to
Zamolodchikov model, in the limit when both sizes of the lattice
tend to infinity. An infinite set of bilinear equations,
describing a distribution of eigenvalues of infinite set of
mutually commuting operators, is derived. The distribution for the
maximal eigenvalues is obtained explicitly. The way to obtain the
excitations is discussed.
\end{abstract}

\maketitle

\section*{Introduction}

Integrability of Zamolodchikov-Baxter three dimensional spin model
\cite{zam,bax-pf} is based on the existence of commutative set of
transfer-matrices $T(\theta_1,\theta_2,\theta_3)$,
\begin{equation}\label{TT}
\biggl[T(\theta_1^{},\theta_2^{},\theta_3^{}),
T(\theta_1^{},\theta_2',\theta_3')\biggr]\;=\;0\;,
\end{equation}
where $\theta_j$ are Zamolodchikov' dihedral angles, and we
understand $T$ as an operator in the vertex formulation
\cite{mss-vertex} of Zamolodchikov-Bazhanov-Baxter model \cite{bb}
with two spin states. In contrast to two-dimensional integrable
models, $T$ is \emph{layer-to-layer} transfer matrix \cite{bs-te},
that means
\begin{itemize}
 \item it is associated with a rectangular lattice with the size
$N\times M$, therefore matrix $T$ has the dimension $2^{NM}\times
2^{NM}$,
 \item \emph{two} parameters $\theta_2$ and
$\theta_3$ are varied in eq. (\ref{TT}).
\end{itemize}
Matrix elements of $R$-matrix of the Zamolodchikov--Baxter model
are not positively defined: it is the obstacle for the decent
interpretation of it as the model of statistical mechanics. The
quantum mechanical interpretation is preferable.

Relation (\ref{TT}) implies the existence of a set of commutative
operators $\{t_{m,n}(\theta_1)\}$ such that
\begin{equation}
\left[t_{m,n}(\theta_1),T(\theta_1,\theta_2,\theta_3)\right]\;=\;0\;\;\;\;\forall\;\;\theta_2,\theta_3,m,n\;,
\end{equation}
i.e. the problem of diagonalization of $T$ for any
$\theta_2,\theta_3$ and the problem of simultaneous
diagonalization of $\{t_{m,n}\}$ are equivalent. In what follows,
we mean a determined definition of $\{t_{m,n}\}$ with $0\leq m\leq
M$ and $0\leq n\leq N$ related to an auxiliary problem of the
model.

It is well known, the Zamolodchikov model and its generalization
-- Bazhanov-Baxter model \cite{bb} -- are related to the
generalized chiral Potts model \cite{gcpm}. The set of
$\{t_{m,n}(\theta_1)\}$ may be produced by the expansion of
transfer matrices of the length $=$ $M$ chain of Lax operators for
cyclic representation of $\mathcal{U}_{q=-1}(\widehat{sl}_N)$
(these $N$ and $M$ are exactly the size of the layer). Another
scheme to produce \emph{the same} set $\{t_{m,n}(\theta_1)\}$ was
proposed in \cite{sergeev-opus,sergeev-aux}. This scheme is the
invariant one from the point of view of $2+1$ dimensional
integrability, in particular the $N\leftrightarrow M$ symmetry is
evident in this scheme. All $t_{m,n}$ are simple polynomials in
the algebra of observables, and besides they are evidently
hermitian (i.e. the model is indeed a model of \emph{quantum
mechanics}). The invariant scheme we mention here as the
\emph{system of interacting spins on a two dimensional lattice},
or as the two dimensional spin lattice.

The eigenvalues of the set $\{t_{m,n}\}$ may be found as a
solution of a system of second order equations. In the language of
auxiliary transfer matrices for generalized chiral Potts, the
system of second order equation is the complete set of fusion
relations for fundamental transfer matrices. The reader may find
the investigation and discussion of the fusion relations and Bethe
Ansatz for Zamolodchikov model for $N=3$ in \cite{kb,bb,bm}. In
the direct $3D$ scheme the whole system of fusion relations is
encoded into a single spectral equation
\cite{sergeev-opus,sergeev-matrix}.

The physical problem is to find the spectrum of all
$t_{m,n}(\theta_1)$ simultaneously. One way to solve this problem
is the nested Bethe Ansatz for the fusion algebra, $M\to\infty$
and finite $N$. Suppose for a moment, somebody has succeeded in
solving the nested Bethe Ansatz equation for arbitrary $N$ (i.e.
in finding the limiting $M\to\infty$ densities of $N-1$ Bethe
Ansatz's distributions of zeros) and then sends $N\to\infty$. From
$2+1$ dimensional point of view such $1+1$ dimensional result
would be related to the limit $N,M\to\infty$ with
$\ds\frac{N}{M}\to 0$, i.e. $N\leftrightarrow M$ symmetry would be
lost -- the model in this approach remains the $1+1$ dimensional
model with infinite symmetry group. This approach would give a
correct answer for a quantity independent on $N/M$.

Contrary to this, the spectral equation in the direct $2+1$ scheme
is initially $N\leftrightarrow M$ invariant. In this paper the
spectral equation is evaluated in the limit
\begin{equation}\label{limit}
N,M\;\to\;\infty\;,\;\;\; \frac{N}{M}\;\to\;\zeta\;\
\end{equation}
where $\zeta$ in the non-singular aspect ratio for the layer. The
main result of this paper is the exact distribution of the largest
eigenvalues (the ground state) $t_{m,n}=f(m,n;\theta_1,\zeta)$ in
the limit (\ref{limit}). The other result is the limiting form of
the spectral equation allowing one to describe (at least
qualitatively) the gap-less excitations of the ground state.

This paper is organized as follows. In sections 1,2 and 3 we
formulate first the system of interacting spins, recall its finite
$N\times M$ -- volume spectral equation and make its leading term
evaluation. Content of the first three sections is a repetition of
\cite{sergeev-opus,sergeev-aux,sergeev-matrix,sergeev-PN,sergeev-pf}.
Next, in the fourth section, we expose some preliminary numerical
results for the spectrum of $t_{m,n}$ and discuss the main idea
for the limiting (\ref{limit}) procedure. In the fifth section
re-write the spectral equations in the thermodynamic limit
$N,M\to\infty$. In the sixth section the qualitative analysis of
the thermodynamical spectral equation is given, the distribution
of the maximal eigenvalues of $t_{m,n}$ is obtained and the
structure of excitations is discussed.

\section{Formulation of the spin lattice system}

All the ways to produce the set $\{t_{m,n}\}$, both via Lax
operators for cyclic representation of
$\mathcal{U}_{q=-1}(\widehat{sl}_N)$ and via $3D$ linear problem
\cite{sergeev-opus}, finally may be reformulated in the following
combinatorial form.

Consider a square lattice with the size $N\times M$ with
periodical boundary conditions -- exactly the layer of (\ref{TT}).
Each vertex $j$ of the lattice may be labelled by the pair of the
indices $j=(n,m)$, $n\in\mathbb{Z}_N$, $m\in\mathbb{Z}_M$. A local
triplet of the Pauli matrices $\sigma^x_{n,m},\sigma^y_{n,m}$ and
$\sigma^z_{n,m}=\ii\,\sigma^x_{n,m}\sigma^y_{n,m}$ is assigned to
each vertex.

Consider a set of non-self-intersecting paths on the periodic
lattice with the following rules of bypassing a vertex and
following factors $\gamma_j$ associated with each variant of
bypassing (note the multiplier $\kappa$ in the third variant):
\begin{center}
\setlength{\unitlength}{0.25mm}
\begin{picture}(340,150)
\put(120,50){\begin{picture}(100,100)
\thinlines\put(50,15){\line(0,1){70}} \put(15,50){\line(1,0){70}}
\put(47,0){\small $n$} \put(0,45){\small $m$}
\put(50,50){\circle*{5}} \thicklines\put(15,15){\line(1,1){35}}
\put(50,50){\line(-1,1){35}}
\put(25,-30){$\gamma_{n,m}=\sigma^y_{n,m}$}
\end{picture}}
\put(240,50){\begin{picture}(100,100)
\thinlines\put(50,15){\line(0,1){70}} \put(15,50){\line(1,0){70}}
\put(47,0){\small $n$} \put(0,45){\small $m$}
\put(50,50){\circle*{5}} \thicklines\put(85,15){\line(-1,1){35}}
\put(50,50){\line(-1,1){35}}
\put(25,-30){$\gamma_{n,m}=\kappa\sigma^z_{n,m}$}
\end{picture}}
\put(0,50){\begin{picture}(100,100)
\thinlines\put(50,15){\line(0,1){70}} \put(15,50){\line(1,0){70}}
\put(47,0){\small $n$} \put(0,45){\small $m$}
\put(50,50){\circle*{5}} \thicklines\put(85,85){\line(-1,-1){35}}
\put(50,50){\line(-1,1){35}}
\put(25,-30){$\gamma_{n,m}=\sigma^x_{n,m}$}
\end{picture}}
\end{picture}
\end{center}
An example of such path for $4\times 4$ lattice is drawn below:
\begin{center}
\setlength{\unitlength}{0.20mm}
\begin{picture}(200,200)
%
%\linethickness{0.1mm}
%
\thinlines \put(25,0){\line(0,1){200}} \put(75,0){\line(0,1){200}}
\put(125,0){\line(0,1){200}} \put(175,0){\line(0,1){200}}
\put(0,25){\line(1,0){200}} \put(0,75){\line(1,0){200}}
\put(0,125){\line(1,0){200}} \put(0,175){\line(1,0){200}}
%
%\put(30,0){\small $1$} \put(80,0){\small $2$} \put(130,0){\small
%$3$} \put(155,0){$\dots$} \put(180,0){\small $N$}
%
%\put(0,30){\small $1$} \put(0,150){$\vdots$} \put(0,80){\small
%$2$} \put(0,130){\small $3$} \put(0,180){\small $M$}
%
\Thicklines \path(50,0)(75,25)(50,50)(75,75)(0,150)
\path(200,150)(175,125)(150,150)(125,125)(100,150)(75,125)(50,150)(75,175)(50,200)
\end{picture}
\end{center}
Any path $\mathcal{P}$ has a homotopy class
$c(\mathcal{P})\;=\;m\mathcal{A}+n\mathcal{B}$, where
$\mathcal{A}$ is the cycle left to right and $\mathcal{B}$ is the
cycle from bottom to top. In the other words, $m$ is the
horizontal winding number and $n$ is the vertical winding number
of the path $\mathcal{P}$. The path at the example above has
$n=m=1$.

For fixed winding numbers $n$ and $m$ let
\begin{equation}
J_{m,n}(\kappa)\;=\;\sum_{\mathcal{P}\;:\;c(\mathcal{P})=m\mathcal{A}+n\mathcal{B}}
\prod_{\mathcal{P}} \gamma_j
\end{equation}
be the sum of the products $\ds \prod_{\mathcal{P}} \gamma_j$ of
$\gamma$-factors along a path $\mathcal{P}$ for all possible paths
with the given winding numbers. In particular, $J_{0,0}\equiv 1$.
The winding numbers of $J_{m,n}$ run $0\leq m \leq M$ and $0\leq
n\leq N$. The reader should distinguish the periodical discrete
coordinates $(n,m)\in(\mathbb{Z}_N,\mathbb{Z}_M)$ of the algebra
of observables and the winding numbers $(m,n)$ labelling the
operators $J$. It is known \cite{sergeev-opus,sergeev-aux},
operators $J_{m,n}$ obey the following exchange relations:
\begin{equation}
J_{m,n} J_{m',n'}\;=\; (-)^{nm'+n'm} \; J_{m',n'} J_{m,n}\;.
\end{equation}
It means, they can be quasi-diagonalized simultaneously:
\begin{equation}
J_{m,n}(\kappa)\;=\;\ii^{nm}\,\left(\sigma^x\right)^m
\left(\sigma^y\right)^n t_{m,n}(\kappa)\;,\;\;\;\left[
t_{m,n}(\kappa), t_{m',n'}(\kappa)\right]\;=\;0\;.
\end{equation}
Auxiliary matrices $\sigma^x$ and $\sigma^y$,
$\sigma^x\sigma^y=-\sigma^y\sigma^x$, belong to the algebra of
observables. They correspond to the homotopy classes
$1\cdot\mathcal{A}+0\cdot\mathcal{B}$ and
$0\cdot\mathcal{A}+1\cdot\mathcal{B}$. Without loss of generality
one may fix
\begin{equation}
\sigma^x\;=\;\prod_n \sigma^x_{n,1}\;\;\;\textrm{and}\;\;\;
\sigma^y\;=\;\prod_m \sigma^y_{1,m}\;.
\end{equation}
The key meaning of the auxiliary matrices is that they represent
one extra degree of freedom of $\{J_{m,n}\}$ with respect to
$\{t_{m,n}\}$. For any of $2^{NM-1}$ eigenstates of $\{t_{m,n}\}$
auxiliary $\sigma^x,\sigma^y$ are usual $2\times 2$ Pauli
matrices.

The set of $J_{m,n}$ (and $\{t_{m,n}\}$ as well) is the set of
``integrals of motion'' for Zamolodchikov model \cite{zam} in its
vertex formulation \cite{mss-vertex}. Namely
\cite{sergeev-opus,sergeev-aux,sergeev-PN}, the layer-to-layer
transfer matrix of Zamolodchikov model
$T(\theta_1,\theta_2,\theta_3)$ commutes with all $J_{m,n}$ for
$\ds\kappa\equiv\tan\frac{\theta_1}{2}$ and arbitrary
$\theta_2,\theta_3$. In this paper we prefer to call $t_{m,n}$ the
\emph{moduli} since in the classical limit they become the moduli
of the classical spectral curve \cite{sergeev-classical}.

The advantage of the present quantum-mechanical formulation is
that if $\kappa$ is real, all $J_{m,n}$ and $t_{m,n}$ are
self-adjoint since the Pauli matrices are self-adjoint, therefore
the model is evidently physical. An eigenstate of the model is
defined by eigenvalues of all $t_{m,n}$ -- one can label the
eigenstates by the corresponding values of $\{t_{m,n}\}$.  Our aim
is to describe all eigenstates.

\section{Finite size spectral equations}

In this section we recall the functional equation for the set of
$t_{m,n}$. For its rigorous derivation see \cite{sergeev-matrix}.

Consider the following generating function:
\begin{equation}\label{generating}
J(x,y)\;=\;\sum_{m=0}^M\;\sum_{n=0}^N  (-)^{n+m+nm} x^m y^n
J_{m,n}\;,
\end{equation}
where $x$ and $y$ are generic complex numbers. In the basis of the
auxiliary $\sigma$-matrices $J(x,y)$ is
\begin{equation}\label{J-via-sigma}
J(x,y)\;=\;t_{0,0}(x,y)-\sigma^x t_{1,0}(x,y) -\sigma^y
t_{0,1}(x,y) - \sigma^z t_{1,1}(x,y)
\end{equation}
where
\begin{equation}\label{Tab}
\begin{array}{ll}
\ds t_{0,0}(x,y)\;=\;\sum_{m,n} x^{2m} y^{2n} t_{2m,2n}\;, & \ds
t_{1,0}(x,y)\;=\;\sum_{m,n} (-)^n x^{2m+1} y^{2n}
t_{2m+1,2n}\;,\\
\\
\ds t_{0,1}(x,y)\;=\;\sum_{m,n} (-)^n x^{2m} y^{2n+1}
t_{2m,2n+1}\;, & \ds t_{1,1}(x,y)\;=\;\sum_{m,n} (-)^{m+n}
x^{2m+1} y^{2n+1} t_{2m+1,2n+1}\;.
\end{array}
\end{equation}
The reader should not be confused by the notation
$t_{\alpha,\beta}(x,y)$ with $\alpha,\beta=0,1$ and the set of
$t_{m,n}$, $0\leq m,n\leq M,N$. It is known
\cite{sergeev-matrix,sergeev-aux,sergeev-PN}, the complete Abelian
algebra of $t_{m,n}$ is generated by the polynomial decomposition
of
\begin{equation}\label{FE}
t_{0,0}(x,y)^2-t_{1,0}(x,y)^2-t_{0,1}(x,y)^2-t_{1,1}(x,y)^2\;=\;F(x^2,y^2)\;,
\end{equation}
where
\begin{equation}\label{Polynom}
F(\lambda^N,\mu^M)\;=\;\prod_{n=0}^{N-1}\prod_{m=0}^{M-1}
(1-\lambda\EXP^{2\pi i n/N} - \mu \EXP^{2\pi i m/M} - \kappa^2
\lambda\mu \EXP^{2\pi i (n/N+m/M)})\;,
\end{equation}
is a polynomial of $\lambda^N=x^2,\mu^M=y^2$:
\begin{equation}\label{13}
F(x^2,y^2)\;=\;\sum_{P=0}^{M}\sum_{Q=0}^{N} x^{2P} y^{2Q}
F_{P,Q}\;.
\end{equation}
As it was mentioned in the introduction, equation (\ref{FE})
encodes the whole fusion algebra of auxiliary transfer matrices
for $U_{q=-1}(\widehat{sl}_N)$, the reader may find its
explanation for e.g. $N=3$ in the appendix.

The right hand side of (\ref{FE}) may be re-written as
\begin{equation}
\sum_{P,Q} x^{2P} y^{2Q} \mathop{\sum\sum}_{m,n}
(-)^{m+n+mn+mQ+nP} t_{m,n}t_{2P-m,2Q-n}\;.
\end{equation}
Equation (\ref{FE}) is the principal solution of the model, in the
same way as the Bethe-ansatz is the principal solution for the
spin chains: the problem of diagonalization of $NM$ $2^{NM}\times
2^{NM}$ matrices $t_{m,n}$ is reduced to a system of $NM$
algebraic equations.
\begin{equation}\label{PQ}
\mathop{\sum\sum}_{m,n} (-)^{m+n+mn+mQ+nP}
t_{m,n}t_{2P-m,2Q-n}\;=\;F_{P,Q}\;.
\end{equation}

\section{The leading term and relation to Zamolodchikov model}

Suppose, no one term is zero in the product (\ref{Polynom}). Then
$F$ in (\ref{FE}) is exponentially big, and one may definitely
conclude \cite{sergeev-pf},
\begin{equation}\label{main-value}
\textrm{Each
of}\;\;\biggl(t_{\alpha,\beta}(x,y)\biggr)^2_{\alpha,\beta=0,1}\;\sim\;
|F(x^2,y^2)|\;\sim\;\EXP^{NM
\mathfrak{g}(\lambda,\mu;\kappa^2)}\;,
\end{equation}
where $x=\lambda^{N/2}$, $y=\mu^{M/2}$, and the integral
\begin{equation}\label{17}
\mathfrak{g}(\lambda,\mu;\kappa^2) \;=\; \lim_{N,M\to\infty}
\frac{1}{NM} \log\left|F(X,Y)\right|\;=\;
\frac{1}{(2\pi)^2}\int\int_0^{2\pi} d\phi d\phi'
\log|1-\lambda\EXP^{\ii\phi}-\mu\EXP^{\ii\phi'}-\kappa^2\lambda\mu\EXP^{\ii(\phi+\phi')}|\;,
\end{equation}
being parameterized by
\begin{equation}
|\lambda|\;=\;\frac{\sin r_2}{\sin r_1}\;,\;\;\;
|\mu|\;=\;\frac{\sin r_3}{\sin r_1}\;,\;\;\;
\kappa^2\;=\;\frac{\sin r_0\sin r_1}{\sin r_2\sin r_3}\;,
\end{equation}
with $r_j$ bounded by
\begin{equation}
r_0+r_1+r_2+r_3=\pi\;\;\; \textrm{and} \;\;\; 0\leq r_1+r_2, \;
r_1+r_3, \; r_2+r_3 <\pi\;,
\end{equation}
has the value \cite{sergeev-pf}
\begin{equation}\label{explicit-f}
\mathfrak{g}(\lambda,\mu;\kappa^2)\;=\; - \log 2\sin r_1 +
\sum_{j=0}^3 \left(\frac{r_j}{\pi}\log 2\sin r_j
+\Phi(r_j)\right)\;,
\end{equation}
where $\Phi(r)$ is the polylogarithm \cite{bax-pf}:
\begin{equation}\label{polylog}
\Phi(r)\;=\;\sum_{m=1}^{\infty}\frac{\sin(2m r)}{2\pi m^2}\;.
\end{equation}
This value is closely related to Baxter's result for the bulk free
energy of the Zamolodchikov model
\cite{bax-pf,sergeev-PN,sergeev-pf}.

Relation (\ref{main-value}) gives the solution  of (\ref{Tab}) for
the mean eigenvalue problem for $t_{\alpha,\beta}(x,y)$: any
eigenvalue of $\biggl(t_{\alpha,\beta}(x,y)\biggr)^2$ has this
leading behavior. The answer (\ref{explicit-f}) is unsatisfactory
from the quantum mechanical point of view, it corresponds to the
asymptotically infinite values of the spectral parameters
$x=\lambda^{N/2}$ and $y=\mu^{M/2}$ and do not clarify the
structure of eigenvalues of the whole set of $t_{m,n}$.

\section{Preliminary evaluation for finite $N,M$}

We started the investigation of (\ref{PQ}) with the numerical
tests for relatively small $N,M$ (up to $N=M=8$) and for simple
choices of $\kappa$ ($\kappa=0,1$).

The principal observation for finite $N,M$ is the following.
Excluding $t_{m,n}$ from (\ref{PQ}) step-by-step, one comes to a
final polynomial equation for a single $t_{m,n}$: such polynomial
equation is exactly the characteristic equation for the operator
$t_{m,n}$. Therefore, the system (\ref{PQ}) and the problem of
direct diagonalization of operators $t_{m,n}$ are equivalent. In
other words, any solution of equations (\ref{PQ}) is indeed an
eigenstate. For this reason we call (\ref{PQ}) the complete
Abelian algebra.

Note in addition the parity property: if a set $\{t_{m,n}\}$
solves the equation (\ref{PQ}), then the set
$\{\widetilde{t}_{m,n}\}$,
\begin{equation}\label{parity}
\widetilde{t}_{2m+\alpha,2n+\beta}=\varepsilon_{\alpha,\beta}
t_{2m+\alpha,2n+\beta}\;,
\end{equation}
where $\alpha,\beta=0,1$ and $\varepsilon_{\alpha,\beta}$ are four
arbitrary signs, solves (\ref{PQ}) as well. This ambiguity
corresponds to the ambiguity of definition of the auxiliary
$\sigma^x,\sigma^y,\sigma^z$.

It is useful to visualize the domain of the indices of
$\{t_{m,n}\}$ as the set $\Pi$ of points $(m,n)$ on the
``momentum'' plane:
\begin{equation}
\Pi\;=\;\{(m,n)\}\;:\;\;\;0\leq n\leq N\;,\;\;\;0\leq m \leq M\;.
\end{equation}
The domain of $F_{P,Q}$ is the same, it is the Newton polygon for
$F(x^2,y^2)$. On the boundary of the rectangular $\Pi$ the
eigenvalues of $t_{m,n}$ as well as the values of $F_{P,Q}$ are
simple. Just putting e.g. $y=0$ in (\ref{J-via-sigma}), one gets
\begin{equation}
t_{0,0}(x,0)^2-t_{1,0}(x,0)^2=(1-x^2)^M\;.
\end{equation}
This equation defines all possible boundary eigenvalues $t_{m,0}$.
Subject of interest is the calculation of $t_{m,n}$ in the middle
of $\Pi$.

Numerical calculations show that for all eigenstates the absolute
values of $t_{m,n}$ as well as the coefficients $F_{m,n}$ grow
significantly when $(m,n)$ goes from the boundary of $\Pi$ to its
middle. One eigenstate (up to the parity equivalence
(\ref{parity})) is strictly separated from all others: absolute
value of any its $t_{m,n}$ is the maximal with respect to values
of the same $t_{m,n}$ for all other eigenstates. We will call it
the ground state.

For a given eigenstate, especially for the ground state, the
values of $t_{m,n}$ are maximal in some $(m,n)=(P_0,Q_0)$ in the
middle of rectangular $\Pi$. In the same point the coefficient
$F_{P_0,Q_0}$ has the maximal absolute value with respect to all
other $F_{P,Q}$. The observed feature of the ground state is that
$\ds\frac{t_{P_0+m,Q_0+n}}{t_{P_0,Q_0}}$ with $|m|$ and $|n|$
being relatively small, depends essentially only on $N/M$ and
$\kappa$ when $N$ and $M$ are big. The same asymptotical
independence of $N,M$ is valid for $\ds
\frac{F_{P_0+m,Q_0+n}}{F_{P_0,Q_0}}$ as well. Since $F_{P,Q}$
takes the maximal value at $(P_0,Q_0)$, expression (\ref{17}) is
the result of a competition between the domain of maximal values
of $F_{P,Q}$ and big (or small) values of $\lambda^{MP}\mu^{NQ}$
accompanying $F_{P,Q}$, sf. (\ref{13}).

Another feature of $\{t_{m,n}\}$ may be mentioned. We observed
that the sets of signs of $\{t_{m,n}\}$ for $(m,n)$ surrounding
$(P_0,Q_0)$ are different (up to (\ref{parity})) for different
eigenstates.

These observations allow us to suggest an idea for evaluation of
(\ref{PQ}). Since both $\{t_{m,n}\}$ and $\{F_{P,Q}\}$ have a
domain of dominance -- the neighborhood of $(P_0,Q_0)$ in the
middle of $\Pi$, far from the boundary  -- one can move to
$(P_0,Q_0)$ and concentrate on its neighborhood. The boundary of
$\Pi$ is far from $(P_0,Q_0)$, and in the limit $N,M\to \infty$
the boundary goes to infinity, so that the domain of the dominance
becomes the open $\mathbb{Z}^2$. It follows for finite $N,M$, in
the neighborhood of $(P_0+m,Q_0+n)$
\begin{equation}
|t_{P_0+m,Q_0+n}|^2\;\sim\;|F_{P_0+m,Q_0+n}|\;\sim\;\EXP^{NM\mathfrak{g}(1,1;\kappa^2)}\;,\;\;\;
|m| \;\textrm{and}\; |n|\;\;\textrm{are small}\;,
\end{equation}
so that singular at $N,M\to\infty$ exponential is just the common
factor for all eigenstates. Cancelling it, one does can evaluate
the spectral equations (\ref{PQ}) in the domain of dominance in
the limit (\ref{limit}). This will be done in the next section.

\section{Spectral equation in the thermodynamic limit}

To rewrite equations (\ref{FE}) or (\ref{PQ}) in the thermodynamic
limit $N,M\to \infty$ with $\ds\zeta=\frac{N}{M}$ being fixed, we
need to introduce several notations.

Define parameters $c$ and $a$ via
\begin{equation}\label{c-and-a}
c\;=\;\cot\frac{a}{2}\;=\;\sqrt{\frac{1+\kappa^2}{3-\kappa^2}}\;\;
\Longleftrightarrow \;\;
\kappa^2\;=\;\frac{\sin\frac{3a}{2}}{\sin\frac{a}{2}}\;.
\end{equation}
At $N,M\to \infty$ the middle point $(P_0,Q_0)$ is defined by
\begin{equation}\label{PQ0}
M\cdot\left(1-\frac{a}{\pi}\right)\;=\;P_0-u_1\;,\;\;\;
N\cdot\left(1-\frac{a}{\pi}\right)\;=\;Q_0-u_2\;
\end{equation}
where $P_0$ and $Q_0$ are \emph{even} integers while $u_1$ and
$u_2$, $-1 < u_1,u_2 \leq 1$, are fractional parts. If $a$ is not
a rational fraction of $\pi$, both $u_1$ and $u_2$ are extra
variables.

Define next the quadratic form parameterized in the terms of $c$
and aspect ratio $\zeta$:
\begin{equation}\label{qf}
\Omega(p,q)\;=\;\frac{\pi}{2}\left(\zeta \frac{1+c^2}{2c}  p^2 +
\frac{1-c^2}{c} pq + \zeta^{-1} \frac{1+c^2}{2c} q^2\right)\;.
\end{equation}
The $N,M\to\infty$ limit of (\ref{PQ}) is based on the following
behavior of the coefficients $F_{P,Q}$:
\begin{equation}\label{fpq}
F_{P_0+p,Q_0+q}\;=\;(-)^{p+q+pq}\;\EXP^{ NM
\mathfrak{g}_0(\kappa^2)}\;\cdot\;\EXP^{-\Omega(p+u_1,q+u_2)}\;\cdot\;
f_0\;\left(1+\frac{f_1+f_2\Omega(p+u_1,q+u_2)}{V}+\dots\right)\;,
\end{equation}
where
\begin{equation}\label{pf}
\mathfrak{g}_0(\kappa^2)\;\equiv\;\mathfrak{g}(1,1;\kappa^2)\;=\;
\left(1-\frac{3a}{2\pi}\right)\log\kappa^2+
3\Phi\left(\frac{a}{2}\right)-\Phi\left(\frac{3a}{2}\right)\;.
\end{equation}
Coefficients $f_0,f_1,f_2$ are some functions of $\kappa^2$,
$\zeta$, $u_1$ and $u_2$ (a sketch derivation of (\ref{fpq}) and
the value of $f_0$ will be given in the appendix).

Define $\tau_{m,n}$ as the fine structure of $t_{m,n}$,
\begin{equation}\label{tau}
t_{P_0+m,Q_0+n}\;=\;\sqrt{f_0}\;\EXP^{\frac{1}{2} NM
\mathfrak{g}_0(\kappa^2)}\; \tau_{m,n}\;.
\end{equation}
Here, according to the idea of the previous section, we have moved
to the middle $(P_0,Q_0)$ and canceled common exponential factor.
Substituting (\ref{fpq}) and (\ref{tau}) into (\ref{PQ}),
cancelling the exponents and taking the limit $N,M\to\infty$, we
come to the following equations for $\tau_{m,n}$,
\begin{equation}\label{FE-tau}
\mathop{\sum\sum}_{m,n\in\mathbb{Z}}
(-)^{m+n+mn}\tau_{p+m,q+n}\tau_{p-m,q-n}\;=\;
\EXP^{-\Omega(p+u_1,q+u_2)}\;.
\end{equation}
The next substitution
\begin{equation}\label{c}
\tau_{m,n}\;=\;c_{m,n}\;\EXP^{-\frac{1}{2}\Omega(m+u_1,n+u_2)}
\end{equation}
transforms (\ref{FE-tau}) into the free from $u_1,u_2$ form:
\begin{equation}\label{FE-c}
\mathop{\sum\sum}_{m,n\in\mathbb{Z}}(-)^{m+n+mn}
\EXP^{-\Omega(n,m)}
c_{p-m,q-n}c_{p+m,q+n}\;=\;1\;\;\;\forall\;\;p,q\;\in\;\mathbb{Z}\;.
\end{equation}
Equations (\ref{FE-tau}) and (\ref{FE-c}) are two forms of
(\ref{FE}) in the thermodynamical limit.

\section{Analysis of (\ref{FE-c})}

For the analysis of (\ref{FE-c}), let us modify it slightly at the
first:
\begin{equation}\label{FE5}
\mathop{\sum\sum}_{m,n\in\mathbb{Z}}(-)^{m+n+mn}
\EXP^{-\beta\Omega(n,m)}
c_{p-m,q-n}c_{p+m,q+n}\;=\;1\;\;\;\forall\;p,q\;\in\;\mathbb{Z}\;,
\end{equation}
where the cut-off parameter $\beta \geq  1$.

Consider for a moment $\zeta=1$. In this case
\begin{equation}
\Omega(m,n)\;=\;\frac{\pi}{4} \left(c^{-1}(m+n)^2 \;+\; c
(m-n)^2\right)\;,
\end{equation}
and we have two small parameters in (\ref{FE5}),
\begin{equation}
Q\;=\;\EXP^{-\beta\pi/4c}\;\;\;\textrm{and}\;\;\;
\widetilde{Q}\;=\;\EXP^{-\beta\pi c/4}\;.
\end{equation}
Equation (\ref{FE5}) may be analyzed in the terms of the
perturbative expansion with respect to $Q, \widetilde{Q}$. The
zero order reads
\begin{equation}
c_{p,q}^2 + o(1) = 1\;\;\;\Rightarrow\;\;\;
c_{p,q}\;=\;\varepsilon_{p,q}\;(1+o(1))\;,
\end{equation}
where $\varepsilon_{p,q}=(\pm)$ is the sign of $c_{p,q}$. In the
first non-trivial order,
\begin{equation}
c_{p,q}\;=\;\varepsilon_{p,q}\;(1+
(\varepsilon_{p+1,q}\varepsilon_{p-1,q}+\varepsilon_{p,q-1}\varepsilon_{p,q+1})Q\widetilde{Q}+\dots)\;.
\end{equation}
This procedure may be continued, the result is a series with
respect to $Q$ and $\widetilde{Q}$,
\begin{equation}\label{series}
c_{p,q}\;=\;\varepsilon_{p,q}\; \biggl( 1 + \sum_{m,n > 0}
\chi_{p,q}^{(m,n)} Q^m \widetilde{Q}^n\biggr)\;.
\end{equation}
with coefficients $\chi^{(m,n)}_{p,q}$ being sums of products of
$\varepsilon_{m,n}$ for $(m,n)$ surrounding $(p,q)$: The first few
nonzero $\chi_{p,q}^{(m,n)}$ with $m+n\leq 4$ are
\begin{equation}
\chi^{(1,1)}_{p,q}\;=\;\varepsilon_{p+1,q}\varepsilon_{p-1,q}+\varepsilon_{p,q-1}\varepsilon_{p,q+1}\;,
\end{equation}
\begin{equation}
\begin{array}{l}
\chi^{(2,2)}_{p,q}\;=\;
\varepsilon_{p,q+1}\varepsilon_{p,q-1}\varepsilon_{p+1,q-1}\varepsilon_{p-1,q-1}
  + \varepsilon_{p+1,q}\varepsilon_{p-1,q}\varepsilon_{p-1,q+1}\varepsilon_{p-1,q-1}
  +\varepsilon_{p,q+1}\varepsilon_{p,q-1}\varepsilon_{p+1,q+1}\varepsilon_{p-1,q+1}\\
  +\varepsilon_{p+1,q}\varepsilon_{p-1,q}\varepsilon_{p,q}\varepsilon_{p-2,q}
  +\varepsilon_{p,q+1}\varepsilon_{p,q-1}\varepsilon_{p,q+2}\varepsilon_{p,q}
  -1+\varepsilon_{p,q+1}\varepsilon_{p,q-1}\varepsilon_{p,q}\varepsilon_{p,q-2}
  -\varepsilon_{p+1,q}\varepsilon_{p-1,q}\varepsilon_{p,q+1}\varepsilon_{p,q-1}\\
  +\varepsilon_{p+1,q}\varepsilon_{p-1,q}\varepsilon_{p+2,q}\varepsilon_{p,q}
  +\varepsilon_{p+1,q}\varepsilon_{p-1,q}\varepsilon_{p+1,q+1}\varepsilon_{p+1,q-1}\;,
\end{array}
\end{equation}
and
\begin{equation}
\chi^{(4,0)}_{p,q}\;=\;\varepsilon_{p-1,q-1}\varepsilon_{p+1,q+1}\;,\;\;\;
\chi^{(0,4)}_{p,q}\;=\;\varepsilon_{p-1,q+1}\varepsilon_{p+1,q-1}\;.
\end{equation}
This procedure may be formulated for $\Omega(m,n)$ with general
$\zeta$ as well.

\begin{conjecture}
If $\beta>1$, all seria  (\ref{series}) converge. Solution of
(\ref{FE5}) is defined uniquely by the distribution of the signs
$\boldsymbol{\varepsilon}\equiv\{\varepsilon_{m,n}\}$.
\end{conjecture}
\noindent Note, due to the parity structure of (\ref{FE5}), any
distribution $\{\varepsilon_{m,n}\}$ is equivalent to
$\{\varepsilon_{m,n}'=\varepsilon_{m,n}^{}(\pm)^{m} (\pm)^{n}\}$,
sf. (\ref{parity}).

The homogeneous distribution $\varepsilon_{m,n}=(+)$ is the
distinguished one since in this case $c_{m,n}=c_0$, expression for
$c_0$ follows from (\ref{FE5}) and matches the series form
(\ref{series}),
\begin{equation}\label{c-value}
c_0\;=\;\left(\sum_{m,n} (-)^{m+n+mn}
\EXP^{-\beta\Omega(m,n)}\right)^{-1/2}\;.
\end{equation}
But if $\beta\to 1$, (\ref{c-value}) diverges as
\begin{equation}\label{ass}
c_0\;\approx\;\frac{1}{\sqrt{(\beta-1)\chi}}
\end{equation}
for some $\chi=\chi(c,\zeta)$. This divergence may be explained by
the $\ds\frac{1}{NM}$ term in (\ref{fpq}):
$\ds\beta=1+\frac{f_2}{NM}$ when $N,M\to\infty$, so that
asymptotically
\begin{equation}
\ds c_0=\sqrt{\frac{NM}{f_2\chi}}\;\sim\;\sqrt{NM}\;.
\end{equation}
The distribution $\varepsilon_{m,n}=(+)$ and
$c_{m,n}=c_0\sim\sqrt{NM}$ is the ground state according to the
numerical tests.

If the signs $\varepsilon_{m,n}$ vary for different $m,n$ (even if
only one sign is opposite to all the others), we have
\begin{conjecture}
The seria (\ref{series}) with inhomogeneous
$\boldsymbol{\varepsilon}$ converge at $\beta=1$.
\end{conjecture}

We can explain $c_0\sim\sqrt{NM}$ in a bit different way. Consider
for instance the following distribution of the signs:
\begin{equation}\label{V-condition}
\varepsilon_{p+m,q+n}\;=\;\left\{\begin{array}{l} \ds
(+)\;\;\textrm{if}\;\; \Omega(m,n) \leq \frac{\pi}{2} V \;,\\
\\ \ds \textrm{randomly}\;\;(\pm)\;\;\textrm{if}\;\;\; \Omega(m,n) > \frac{\pi}{2} V
\end{array}
\right.
\end{equation}
In this case a \emph{very rough} estimation gives
\begin{equation}\label{bulk}
c_{p,q}\;\sim\;\sqrt{V}\;.
\end{equation}
Thus the finite-volume domain of positive signs on the infinite
lattice is effectively equivalent to finite lattice, and the strip
$\ds\Omega(m,n)\sim\frac{\pi}{2} V$ plays the r\^ole of an
effective boundary.

The homogeneous distribution $\varepsilon_{m,n}=(+)$ in a big
volume $V$ gives evidently the maximal eigenvalues of the quantum
mechanical model, any variation of the signs gives an excitation
of the spectrum. A distribution of the signs
$\varepsilon_{m_1,n_1}=\varepsilon_{m_2,n_2}=...\varepsilon_{m_k,n_k}=(-)$
with  $(m_1,n_1)...(m_k,n_k)$ inside $V$ and with all other
$\varepsilon_{m,n}=(+)$ inside $V$, is a candidate for a
$k$-particles state.

One particle state, $\varepsilon_{m_1,n_1}=(-)$ with all other
$\varepsilon_{m,n}=(+)$ inside $V$, is described asymptotically by
two continuous parameters $\ds
(\mu,\nu)=\left(\frac{m_1}{\sqrt{V}},\frac{n_1}{\sqrt{V}}\right)$.
We expect a ``dispersion relation'' in the form $\ds
\frac{\tau_{m,n}}{\sqrt{V}} = \textrm{a smooth function of }
(\mu,\nu)$. The model evidently is gap-less.

The behavior (\ref{bulk}) allows one to suggest a candidate for
the Hamiltonian of the system:
\begin{equation}\label{hamiltonian-1}
H\;=\;-\sum_{m,n} \tau_{m,n}^2\;\equiv \;-\sum_{m,n} c_{m,n}^2
\EXP^{-\Omega(m,n)}\;,
\end{equation}
sf. (\ref{c}). At the ground state $H\;\approx\; -h_0 V$, i.e. one
can talk about the density energy $-h_0$ of the ground state, and
the spectrum of $H$ describes bound states $\ds -h_0\leq
\frac{H}{V}<0$.

From the alternative point of view, one may consider the
Hamiltonian
\begin{equation}\label{hamiltonian-2}
H'\;=\;-H\;.
\end{equation}
For this Hamiltonian, the ground state corresponds to a random
distribution of the signs -- we can say nothing about it.
Excitations are the islands of constant signs in the sea of random
ones, and its maximal value is described by the finite energy
density $+h_0$.

\section{Discussion}

The main results of this paper are the following. The ground state
distribution of the moduli
\begin{equation}
\tau_{p,q}\;=\;c_{p,q}\EXP^{-\frac{1}{2}\Omega(p,q)}
\end{equation}
related to the moduli $t_{m,n}$ by (\ref{tau}), is given by
\begin{equation}
\tau_{p,q}\;=\;c_0\;\EXP^{-\frac{1}{2}\Omega(p,q)}\;,\;\;\;c_0\;\sim\;\sqrt{V}
\end{equation}
where $V$ is the volume of the system. Any excited state is
uniquely defined by a distribution of the signs
$\varepsilon_{p,q}=\textrm{sign of } c_{p,q}$, the set of
$c_{p,q}$ is the solution of
\begin{equation}\label{53}
\sum_{m,n}(-)^{m+n+mn} \EXP^{-\Omega(n,m)}
c_{p-m,q-n}c_{p+m,q+n}\;=\;1\;.
\end{equation}
The various problems of our approach are to be mentioned. At the
first, we are still unable to evaluate (\ref{53}) explicitly for
non-periodical distribution of signs. Even $f_2$ in the asymptotic
(\ref{fpq}) or more exact estimation of (\ref{bulk}) are not
known. Also it is not known yet how to express the momenta,
corresponding to two orthogonal shifts of the periodical lattice,
in the terms of $\{t_{m,n}\}$. Without the physical momenta, one
hardly can interpret physically the dispersion relation
conjectured above. The third problem is the calculation of the
spectrum of $T(\theta_1,\theta_2,\theta_3)$. It is still unclear
how the transfer matrix of Zamolodchikov-Baxter model is related
to $\{t_{m,n}\}$ for generic $N,M$. Some discussion repeating
\cite{sergeev-pf} is given in appendix.

\noindent \textbf{Acknowledgements} The author should like to
thank V. Bazhanov, V. Mangazeev, M. Batchelor, X-W. Guan and all
the Mathematical Physics group of the Department of Theoretical
Physics of RSPhysSE for fruitful discussions.

\appendix

\section{$sl_3$ fusion algebra}

Let us demonstrate how the equations (\ref{FE},\ref{Polynom})
generate the fusion algebra of auxiliary transfer matrices. Choose
the particular value $N=3$. The series (\ref{J-via-sigma}) may be
rewritten as the four-term sum
\begin{equation}
\begin{array}{l}
\ds J(x,y)\;=\;\sum_{n=0}^N\sum_{m=0}^M(-\ii)^{mn} (-x\sigma^x)^m
(-y\sigma^y)^n t_{m,n}\;=\;\\
\\
\ds \left(\sum_{m=0}^M (-x\sigma^x)^m t_{m,0}\right) -
\left(\sum_{m=0}^M (\ii x \sigma^x)^m t_{m,1}\right) y\sigma^y +
\left(\sum_{m=0}^M (x\sigma^x)^m t_{m,2}\right) y^2 -
\left(\sum_{m=0}^M (-\ii x\sigma^x)^m t_{m,3}\right)
y^3\sigma^y\\
\\
\ds \equiv\; t_0(x\sigma^x) \;-\; t_1(x\sigma^x)\; y\sigma^y \;+\;
t_2(x\sigma^x)\; y^2 \;-\; t_3(x\sigma^x)\; y^3\sigma^y\;.
\end{array}
\end{equation}
One may show combinatorially \cite{sergeev-aux}, $t_k(x)$ is the
$\mathcal{U}_{q=-1}(\widehat{sl}_N)$ transfer matrix for the Lax
operators with the cyclic representation in the quantum space and
the fundamental representation $\pi_k$ in the auxiliary space. It
is supposed, $\pi_0$ and $\pi_N$ are the scalar representations,
$\pi_1$ is the vector representation etc. In $N=3$ case $\pi_2$ is
the co-vector representation.

Decomposition of $F(x^2,y^2)$ with respect to $y^2$ is following
($\ds \omega=\EXP^{2\pi\ii/3}$ and $\lambda^3=x^2$):
\begin{equation}
F(x^2,y^2)\;=\;\prod_{n=0}^2 \biggl(
(1-\lambda\omega^n)^M-y^2(1+\kappa^2\lambda\omega^n)^M\biggr)\;=\;
A(x^2)-B(x^2)y^2+C(x^2)y^4-D(x^2)y^6\;,
\end{equation}
where
\begin{equation}
\begin{array}{l}
\ds A(x^2)\;=\;(1-x^2)^M\;,\;\;\;\ds D(x^2)\;=\;(1+\kappa^6
x^2)^M\;,\\
\\
\ds
B(x^2)\;=\;(1-x^2)^M\left(\left(\frac{1+\kappa^2\lambda}{1-\lambda}\right)^M+
\left(\frac{1+\kappa^2\lambda\omega}{1-\lambda\omega}\right)^M+
\left(\frac{1+\kappa^2\lambda\omega^2}{1-\lambda\omega^2}\right)^M\right)\;,\\
\\
\ds C(x^2)\;=\;(1+\kappa^6
x^2)^M\left(\left(\frac{1-\lambda}{1+\kappa^2\lambda}\right)^M+
\left(\frac{1-\lambda\omega}{1+\kappa^2\lambda\omega}\right)^M+
\left(\frac{1-\lambda\omega^2}{1+\kappa^2\lambda\omega^2}\right)^M\right)\;.
\end{array}
\end{equation}
Equating now (\ref{FE}) in all orders of $y^2$, one comes at $y^0$
and $y^6$ to
\begin{equation}
t_0(x)t_0(-x)\;=\;(1-x^2)^M\;,\;\;\;t_3(x)t_3(-x)\;=\;(1+\kappa^6x^2)^M\;.
\end{equation}
For the generalized chiral Potts model the choice is prescribed:
\begin{equation}\label{cpm-choice}
t_0(x)=(1-x)^M\;,\;\;\;\; t_3(x)\;=\;(1+\ii\kappa^3x)\;.
\end{equation}
The orders $y^2$ and $y^4$ give
\begin{equation}\label{fusion}
\begin{array}{l}
\ds t_1(x)t_1(-x) \;=\; t_0(-x) t_2(x) + t_0(x) t_2(-x) +
B(x^2)\;,\\
\\
\ds t_2(x) t_2(-x) \;=\; t_3(-x) t_1(x) + t_3(x) t_1(-x) +
C(x^2)\;.
\end{array}
\end{equation}
Relations (\ref{fusion}) with (\ref{cpm-choice}) are exactly the
fusion algebra for $sl_3$, \cite{kb,bb,bm}.

\section{Asymptotic of $F_{P,Q}$}

Let us discuss briefly the derivation of (\ref{fpq}). Taking into
account (\ref{explicit-f}), one may use the saddle point method
for the estimation of $F_{P,Q}$. Basically,
\begin{equation}
F_{P,Q}\;=\;\frac{1}{(2\pi \ii)^2} \oint\oint \frac{dX}{X}
\frac{dY}{Y} \frac{F(X,Y)}{X^PY^Q}\;.
\end{equation}
Let
\begin{equation}
\alpha_p\;=\;\frac{P\pi}{M}\;,\;\;\;
\alpha_q\;=\;\frac{Q\pi}{N}\;.
\end{equation}
Then
\begin{equation}
\log\left(\frac{F(X,Y)}{X^P Y^Q}\right)\;\sim\;NM\;\biggl(
\mathfrak{g}(\lambda,\mu;\kappa^2)-\frac{\alpha_p}{\pi}\log\lambda
-\frac{\alpha_q}{\pi}\log\mu\biggr)
\end{equation}
It has the extremum (minimum) with respect to $\lambda,\mu$
($\kappa^2$ being fixed) at \footnote{In details, $\ds
\lambda\frac{\partial
\mathfrak{g}}{\partial\lambda}\;=\;\frac{r_0+r_2}{\pi}$,  $\ds
\mu\frac{\partial\mathfrak{g}}{\partial\mu}\;=\;\frac{r_0+r_3}{\pi}$,
$\ds \kappa^2\frac{\partial\mathfrak{g}}
{\partial\kappa^2}\;=\;\frac{r_0}{\pi}$.}
\begin{equation}
r_0+r_2\;=\;\alpha_p\;,\;\;\; r_0+r_3\;=\;\alpha_q\;.
\end{equation}
The extremum value of $\ds
\mathfrak{g}(\lambda,\mu;\kappa^2)-\frac{\alpha_p}{\pi}\log\lambda
-\frac{\alpha_q}{\pi}\log\mu$ is
\begin{equation}
g(\alpha_p,\alpha_q;\kappa^2)\;=\; \frac{r_0}{\pi}\log \kappa^2 +
\sum_{j=0}^3 \Phi(r_j)
\end{equation}
where the numbers $r_j$ are to be calculated via
\begin{equation}
r_0\;=\;\pi-\frac{a_1+a_2+a_3}{2}\;,\;\;\;
r_1\;=\;\frac{a_2+a_3-a_1}{2}\;,\;\;\;
r_2\;=\;\frac{a_3+a_1-a_2}{2}\;,\;\;\;
r_3\;=\;\frac{a_1+a_2-a_3}{2}\;,
\end{equation}
and
\begin{equation}
a_2\;=\;\pi-\alpha_p\;,\;\;\; a_3\;=\;\pi-\alpha_q\;,\;\;\;
a_1\;=\;\arccos \left(\cos a_2\cos a_3
+\frac{\kappa^2-1}{\kappa^2+1} \sin a_2\sin a_3\right)\;.
\end{equation}
The last equality is the solution of $\ds\kappa^2=\frac{\sin
r_0\sin r_1}{\sin r_2\sin r_3}$ with respect to $a_1$. Therefore
asymptotically
\begin{equation}\label{fpq-1}
F_{P,Q}\;=\;(-)^{P+Q+PQ}\;\cdot\; f_0 \;\cdot\;
\left(1+\frac{F'}{NM}+...\right)\;\cdot\; \EXP^{NM
g(\alpha_p,\alpha_q;\kappa^2)}\;.
\end{equation}
Function $g(\alpha_q,\alpha_q;\kappa^2)$ has the maximum near
$\alpha_p=\alpha_q=\pi-a$, where $a$ is defined by
(\ref{c-and-a}), and
\begin{equation}
g(\alpha_p,\alpha_q;\kappa^2)\;=\;\mathfrak{g}_0(\kappa^2)\;-\;
\frac{1+c^2}{4\pi c}(\delta\alpha_p^2+\delta\alpha_q^2) \;-\;
\frac{1-c^2}{2\pi c} \delta\alpha_p\delta\alpha_q\;,
\end{equation}
where $\mathfrak{g}_0(\kappa^2)$ is given by (\ref{pf}). Let
further even integers $P_0,Q_0$ and real numbers $u_1,u_2$ are
defined by (\ref{PQ0}). Then
\begin{equation}
\delta\alpha_p\;=\;\frac{\pi}{M}(p+u_1)\;,\;\;\;
\delta\alpha_q\;=\;\frac{\pi}{N}(q+u_2)\;.
\end{equation}
Therefore, the leading term of (\ref{fpq-1}) is
\begin{equation}\label{fpq-2}
F_{P_0+p,Q_0+q}\;=\;(-)^{p+q+pq} \;\cdot\;f_0\;\cdot\; \EXP^{NM
\mathfrak{g}_0(\kappa^2)-\Omega(p+u_1,q+u_2)}\;,
\end{equation}
where the quadratic form is given by (\ref{qf}).

The next order in (\ref{fpq-1}), $F'=f_1+f_2\Omega(p+u_1,q+u_2)$,
is the result of numerical tests.

\section{Theta-functions}

In the limit $M,N\to\infty$ the polynomial $F(X,Y)$ as well as the
eigenstates of $t_{\alpha,\beta}(x,y)$ for periodical distribution
of the signs $\varepsilon_{m,n}$ become the theta-functions. In
particular, equations (\ref{FE-tau}) may be re-written in a
theta-functions-like form:
\begin{equation}\label{FE4}
\begin{array}{l}
\ds \biggl(\sum x^{2m}y^{2n}\tau_{2m,2n}\biggr)^2 - \biggl(\sum
(-)^nx^{2m+1}y^{2n}\tau_{2m+1,2n}\biggr)^2  - \biggl(\sum
(-)^mx^{2m}y^{2n+1}\tau_{2m,2n+1}\biggr)^2 \\
\\
\ds - \biggl(\sum
(-)^{n+m}x^{2m+1}y^{2n+1}\tau_{2m+1,2n+1}\biggr)^2  = \sum_{p,q}
(-)^{p+q+pq} \EXP^{-\Omega(p+u_1,q+u_2)} x^{2p} y^{2q}
\end{array}
\end{equation}
Let us re-define ($x=\EXP^{\ii\pi z_1}$, $y=\EXP^{\ii\pi z_2}$).
Then the theta-function-like seria
\begin{equation}\label{newT}
\tau_{\alpha,\beta}(z_1,z_2)\;=\;\sum_{m,n\in\mathbb{Z}}
(-)^{\alpha n+\beta m} \tau_{2m+\alpha,2n+\beta} \EXP^{\ii\pi
(2m+\alpha) z_1 +\ii\pi (2n+\beta) nz_2}.
\end{equation}
stand for the transfer matrices.

It is helpful to discuss some properties of theta-functions. Let
\begin{equation}\label{def-theta}
\Theta_{u_1,u_2}^{(\beta)}(z_1,z_2)\;=\; \sum_{p,q}
\EXP^{-\beta\Omega(p+u_1,q+u_2)+2\pi\ii pz_1 + 2\pi\ii qz_2}
\end{equation}
for our particular quadratic form $\Omega$ (\ref{qf}). It has the
general Jacobi transform property:
\begin{equation}\label{Jacoby}
\Theta_{u_1,u_2}^{(\beta)}(z_1,z_2)\;=\; \frac{2}{\beta}\;
\EXP^{-2\pi\ii (z_1u_1+z_2u_2)}\;
\Theta_{z_2,-z_1}^{(4/\beta)}(-u_2,u_1)\;.
\end{equation}
The other $\theta$-function, related to $F$, is
\begin{equation}\label{def-pi}
\digamma_{u_1,u_2}(z_1,z_2)\;=\;\sum_{p,q}\;
(-)^{p+q+pq}\EXP^{-\Omega(p+u_1,q+u_2)+2\pi\ii z_1 p + 2\pi\ii z_2
q}\;.
\end{equation}
One can easily see,
\begin{equation}
\begin{array}{l}
\digamma_{u_1,u_2}(z_1,z_2) \;=\; {\ds \frac{1}{2}}
\left(\Theta^{(1)}_{u_1,u_2}(z_1+\frac{1}{2},z_2+ \frac{1}{2}) +
\Theta^{(1)}_{u_1,u_2}(z_1+\frac{1}{2},z_2) +
\Theta^{(1)}_{u_1,u_2}(z_1,z_2+\frac{1}{2}) -
\Theta^{(1)}_{u_1,u_2}(z_1,z_2)\right)\\
\\
=\left(2\Theta^{(4)}_{u_1/2,u_2/2}(2z_1,2z_2) -
\Theta^{(1)}_{u_1,u_2}(z_1,z_2)\right)\;.
\end{array}
\end{equation}
For the case $u_1=u_2=0$ the polynomial identity $
F_{2N,2M}(x^2,y^2) = F_{N,M}(x,y) F_{N,M}(-x,y) F_{N,M}(x,-y)
F_{N,M}(-x,-y)$ provides
\begin{equation}
f_0\digamma_{0,0}(z_1,z_2)\;=\; f_0^4
\digamma_{0,0}(\frac{z_1}{2},\frac{z_2}{2})
\digamma_{0,0}(\frac{z_1+1}{2},\frac{z_2}{2})
\digamma_{0,0}(\frac{z_1}{2},\frac{z_2+1}{2})
\digamma_{0,0}(\frac{z_1+1}{2},\frac{z_2+1}{2})\;.
\end{equation}
The limit $z_1,z_2\to 0$ gives $f_0$ for (\ref{fpq}):
\begin{equation}
f_0\;=\;\sqrt[3]{\frac{4}{\digamma_{0,0}(\frac{1}{2},0)
\digamma_{0,0}(0,\frac{1}{2})
\digamma_{0,0}(\frac{1}{2},\frac{1}{2})}}\;.
\end{equation}
As well, the value of $\chi$ for (\ref{ass}) follows from
\begin{equation}
\sum_{m,n} (-)^{m+n+mn}\EXP^{-\beta
\Omega(m,n)}\;=\;\digamma^{(\beta)}_{0,0}(0,0)\;=\;
\frac{1}{\beta}\Theta_{0,0}^{(1/\beta)}-\Theta_{0,0}^{(\beta)}\;\approx\;(1-\beta)\chi\;
\end{equation}
at $\beta\to 1$ with $\ds \chi =
\Theta_{0,0}^{(1)}+2\frac{\partial\Theta_{0,0}^{(\beta)}}{\partial\beta}|_{\beta=1}$.

\section{Examples of periodical distribution}

Here we give an example is a periodical distribution of the signs.
Let
\begin{equation}
\varepsilon_{2m+\alpha,2n+\beta}\;=\;\varepsilon_{\alpha,\beta}\;\EXP^{\ii\pi
(um+vn)}
\end{equation}
with $u,v=0\;\;\textrm{or}\;\;1$. Periodicity of
$\varepsilon_{m,n}$ provides the periodicity of the series
expansions (\ref{series}), and therefore
\begin{equation}
c_{2m+\alpha,2n+\beta}\;=\;\varepsilon_{2m+\alpha,2n+\beta}\;c_{\alpha,\beta}\;.
\end{equation}
Equation (\ref{FE5}) gives
\begin{equation}
c_{\alpha,\beta}^2\Theta^{(4\beta)}_{0,0}-c_{1-\alpha,\beta}^2\EXP^{\ii\pi
u}\Theta_{\frac{1}{2},0}^{(4\beta)} - c_{\alpha,1-\beta}^2
\EXP^{\ii\pi v} \Theta_{0,\frac{1}{2}}^{(4\beta)} -
c_{1-\alpha,1-\beta}^2 \EXP^{\ii\pi(u+v)}
\Theta_{\frac{1}{2},\frac{1}{2}}^{(4\beta)}\;=\;1
\end{equation}
for all four choices of $(\alpha,\beta)$, its solution is
$c_{0,0}^2=c_{1,0}^2=c_{0,1}^2=c_{1,1}^2$ (it follows as well from
the careful analysis of the structure of
$\varepsilon_{m,n}$-products in (\ref{series})), so that
\begin{equation}
c_{2m+\alpha,2n+\beta}\;=\;\varepsilon_{\alpha,\beta}\;\EXP^{\ii\pi
(um+vn)}\; \left(\Theta_{0,0}^{(4\beta)}-\EXP^{\ii\pi
u}\Theta_{\frac{1}{2},0}^{(4\beta)}-\EXP^{\ii\pi v}
\Theta_{0,\frac{1}{2}}^{(4\beta)} - \EXP^{\ii\pi(u+v)}
\Theta_{\frac{1}{2},\frac{1}{2}}^{(4\beta)}\right)^{-1/2}\;.
\end{equation}

\section{Transfer matrix of Zamolodchikov-Bazhanov-Baxter model}

In the last section we would like to describe the relation between
(\ref{explicit-f}) and Baxter's free energy for Zamolodchikov's
model. We will refer to \cite{sergeev-PN}, where the inhomogeneous
model was considered and \emph{divisor} parameterization was used.
Equations (231) in \cite{sergeev-PN} look like
\begin{equation}\label{JT}
J(X)\,\cdot\, \mathbf{T} \;=\; \mathbf{T}\,\cdot\, J(X')\;=\;0\;.
\end{equation}
Here $J(X)$ and $J(X')$ are generating functions
(\ref{generating}), operator $\mathbf{T}$ is a modified transfer
matrix for Zamolodchikov-Bazhanov-Baxter model (in general, the
Pauli matrices may be replaced by the Weyl algebra generators at
root of unity). It follows from (\ref{JT}), $\mathbf{T}$ up to a
normalization is the product of algebraic supplements of $J(X)$
and $J(X')$.

In our particular case, $J(X)$, $J(X')$ and $\mathbf{T}$ after the
quasi-diagonalization are $2\times 2$ matrices (in the basis of
the Pauli matrices). Transfer-matrix of Zamolodchikov's model
$T(\theta_1,\theta_2,\theta_3)$, mentioned in the Introduction, is
the trace of $\mathbf{T}$:
\begin{equation}
T\;=\;\textrm{Trace}_{2\times 2} \mathbf{T}\;.
\end{equation}
Generating functions $J(X)$ and $J(X')$ stand for
$J(\lambda(X)^{N/2},\mu(X)^{M/2};\kappa^2)$ and
$J(\lambda(X')^{N/2},\mu(X')^{M/2};\kappa^2)$ in the present
notations, where
\begin{equation}
\kappa^2\;=\;\tan^2\frac{\theta_1}{2}\;=\;\frac{\sin\beta_2\sin\beta_3}{\sin\beta_0\sin\beta_1}
\end{equation}
is the $\kappa$-parameter in both $J(X)$ and $J(X')$, and explicit
evaluations for $\lambda$ and $\mu$ from \cite{sergeev-PN} to the
terms of linear excesses $\beta_j$ give
\begin{equation}
\lambda(X)\;=\;\EXP^{-\ii(\beta_1+\beta_2)}\frac{\sin\beta_0}{\sin\beta_3}\;,\;\;\;
\mu(X)\;=\;\EXP^{\ii(\beta_0+\beta_2)}\frac{\sin\beta_1}{\sin\beta_3}\;,
\end{equation}
and
\begin{equation}
\lambda(X')\;=\;\EXP^{\ii(\beta_0+\beta_3)}
\frac{\sin\beta_1}{\sin\beta_2}\;,\;\;\;
\mu(X')\;=\;\EXP^{-\ii(\beta_1+\beta_3)}\frac{\sin\beta_0}{\sin\beta_2}\;.
\end{equation}
It gives us the identification $\{r_j\}=\{\textrm{a permutation
of}\;\; \beta_j\}$ and relates (\ref{explicit-f}) to Baxter's
answer for the partition function per site $k$:
\begin{equation}\label{baxter-pf}
\log k\;=\;\textrm{normalization}\;+\;\sum_{j=0}^3
\left(\frac{\beta_j}{2\pi}\log 2\sin\beta_j
+\Phi(\beta_j)\right)\;.
\end{equation}
The reader may see the discrepancy, $\ds \frac{\beta_j}{2\pi}\log
2\sin\beta_j$ in (\ref{baxter-pf}) and $\ds
\frac{\beta_j}{\pi}\log 2\sin\beta_j$ in (\ref{explicit-f}), it
means that the normalization is not trivial -- it comes from a
certain variational principle.


\begin{thebibliography}{99}

\bibitem{bax-pf}
R. J. Baxter 1983 On Zamolodchikov's solution of the tetrahedron
equation \emph{Commun. Math. Phys.} \textbf{88} 185-205

R. J. Baxter 1984 Partition function of the three-dimensional
Zamolodchikov model \emph{Phys. Rev. Lett.} \textbf{53} 1795

R. J. Baxter 1986 The Yang-Baxter equations and the Zamolodchikov
model \emph{Physica} \textbf{18D} 321-347


\bibitem{zam}
A. B.  Zamolodchikov, ``Tetrahedron equations and integrable
systems in three dimensions'', \emph{JETP} \textbf{79} (1980)
641-664 (in russian)

A. B. Zamolodchikov, ``Tetrahedron equations and the relativistic
S matrix of straight strings in 2+1 dimensions'', \emph{Commun.
Math. Phys.} \textbf{79} (1981) 489-505

\bibitem{mss-vertex}
S. M. Sergeev, V. V. Mangazeev and Yu. G. Stroganov, ``Vertex
reformulation of the Bazhanov -- Baxter model'', \emph{J. Stat.
Phys.} \textbf{82} (1996) 31-50

\bibitem{bb}
V. V. Bazhanov and R. J. Baxter, ``New solvable lattice models in
three dimensions'', \emph{J. Stat. Phys.} \textbf{69} (1992)
453-485

\bibitem{bs-te}
V. Bazhanov and Yu. Stroganov, ``Conditions of commutativity of
transfer-matrices on a multidimensional lattice'', \emph{Theor.
Math. Phys.} \textbf{52} (1982) 685-691

\bibitem{gcpm}
V. V.  Bazhanov, R. M.  Kashaev, V. V. Mangazeev and Yu. G.
Stroganov, ``$Z_N{\otimes {n-1}}$ generalization of the chiral
Potts model'', \emph{Comm. Math. Phys.} \textbf{138} (1991)
393--408

\bibitem{sergeev-opus}
S. Sergeev, ``Quantum {$2+1$} evolution model'', \emph{J. Phys. A:
Math. Gen.} \textbf{32} (1999) 5693--5714

\bibitem{sergeev-aux}
S. M. Sergeev, ``Auxiliary transfer matrices for
three-dimensional integrable models'', \emph{Theoretical and
Mathematical Physics} \textbf{124} (2000) 391--409

\bibitem{kb}
V. V. Bazhanov and R. M. Kashaev, ``Cyclic $L$-operator related
with a $3$-state $R$-matrix'', \emph{Comm. Math. Phys.}
\textbf{136} (1991) 607--623

\bibitem{bm}
H. E. Boos and V. V. Mangazeev, ``Functional relations and nested
Bethe ansatz for $sl(3)$ chiral Potts model at $q^2=-1$'',
\emph{J. Phys. A: Math. Gen} \textbf{32} (1999) 3041-3054

H. E. Boos and V. V. Mangazeev, ``Bethe ansatz for the three-layer
Zamolodchikov model'', \emph{J. Phys. A: Math. Gen.} \textbf{32}
(1999) 5285-5298

H. E. Boos and V. V. Mangazeev, ``Some exact results for the
three-layer Zamolodchikov model'', \emph{Nucl. Phys. B}
\textbf{592} (2001) 597-626

\bibitem{sergeev-matrix}
S. M. Sergeev, ``Coefficient Matrices of a Quantum
Discrete Auxiliary Linear Problem'', \emph{Journal of Mathematical
Sciences} \textbf{115}(1) (2003) 2049-2057

\bibitem{sergeev-PN}
S. Sergeev, ``Quantum integrable models in discrete 2+1
dimensional space-time: auxiliary linear problem on a lattice,
zero curvature representation, isospectral deformation of the
Zamolodchikov-Bazhanov-Baxter model'', the review accepted in
\emph{Particles and Nuclei} (2004).


\bibitem{sergeev-pf}
S. M. Sergeev, ``Evidence for a phase transition in three
dimensional lattice models'', \emph{Theoretical and Mathematical
Physics} \textbf{138} (2004) 310-321

\bibitem{sergeev-classical}
S. M. Sergeev, ``On exact solution of a classical 3{D} integrable
model'', \emph{J. Nonlinear Math. Phys.} \textbf{1} (2000) 57--72

\end{thebibliography}
\end{document}